\documentclass[preprint, superscriptaddress, draft, amssymb, floats, byrevetex]{revtex4}
\usepackage{latexsym}
\usepackage{graphics}

\newcommand{\bmat}{\left(\begin{array}}
\newcommand{\emat}{\end{array}\right)}
\newcommand{\beq}{\begin{equation}}
\newcommand{\eeq}{\end{equation}}

\newcommand{\drawsquare}[2]{\hbox{%
\rule{#2pt}{#1pt}\hskip-#2pt
\rule{#1pt}{#2pt}\hskip-#1pt
\rule[#1pt]{#1pt}{#2pt}}\rule[#1pt]{#2pt}{#2pt}\hskip-#2pt
\rule{#2pt}{#1pt}}

\newcommand{\fund}{\raisebox{-.5pt}{\drawsquare{6.5}{0.4}}}
\newcommand{\Ysymm}{\raisebox{-.5pt}{\drawsquare{6.5}{0.4}}\hskip-0.4pt%
        \raisebox{-.5pt}{\drawsquare{6.5}{0.4}}}
\newcommand{\Yasymm}{\raisebox{-3.5pt}{\drawsquare{6.5}{0.4}}\hskip-6.9pt%
        \raisebox{3pt}{\drawsquare{6.5}{0.4}}}
\newcommand{\antifund}{\overline{\fund}}

\def\yzero{\smash{\hbox{$y\kern-4pt\raise1pt\hbox{${}^\circ$}$}}}

\def\-{\hphantom{-}}

\def\s2{\frac{1}{\sqrt2}}

\def\beq{\begin{equation}}
\def\eeq{\end{equation}}
\def\beqa{\begin{eqnarray}}
\def\eeqa{\end{eqnarray}}

\def\IF{\relax{\rm I\kern-.18em F}}
\def\II{\relax{\rm I\kern-.18em I}}
\def\IP{\relax{\rm I\kern-.18em P}}

\def\Dsl{\,\raise.15ex\hbox{/}\mkern-13.5mu D} 

\def\IC{\bf C}
\def\IZ{\bf Z}

\def\z2z2{$\IC^3/(\IZ_2\times\IZ_2)$}

\def\s{\sigma}

\def\z{\zeta}

\def\bo{{\raise-.3ex\hbox{\large$\Box$}}}               
\def\face{{\raise.2ex\hbox{$\displaystyle \bigodot$}\mskip-2.2mu \llap {$\ddot
        \smile$}}}                                      

\def\leftrightarrowfill{$\mathsurround=0pt \mathord\leftarrow \mkern-6mu
        \cleaders\hbox{$\mkern-2mu \mathord- \mkern-2mu$}\hfill
        \mkern-6mu \mathord\rightarrow$}       
\def\dvec#1{\vbox{\ialign{##\crcr
        \leftrightarrowfill\crcr\noalign{\kern-1pt\nointerlineskip}
        $\hfil\displaystyle{#1}\hfil$\crcr}}}           

\def\beq{\begin{equation}}
\def\eeq{\end{equation}}

\def\beqx{\begin{displaymath}}
\def\eeqx{\end{displaymath}}

\def\beqa{\begin{eqnarray}}
\def\eeqa{\end{eqnarray}}

\begin{document}

\DeclareGraphicsExtensions{.jpg,.pdf,.mps,.png}

\title{
\normalsize \mbox{ }\hspace{\fill}
\begin{minipage}{12 cm}
{\tt ~~~~~~~~~~~~~~~~~~ UPR-1101-T, hep-th/0501041}{\hfill}
\end{minipage}\\[5ex]
{\large\bf Standard-like Models as Type IIB  Flux  Vacua
\\[1ex]}}
\date{\today}
\author{Mirjam Cveti\v c}
\affiliation{ Department of Physics and Astronomy, University of Pennsylvania, \\
Philadelphia, PA 19104, USA}
\author{Tianjun Li}
\affiliation{School of Natural Science, Institute for Advanced Study,  \\
             Einstein Drive, Princeton, NJ 08540, USA}
\author{Tao Liu}
\affiliation{ Department of Physics and Astronomy, University of Pennsylvania, \\
Philadelphia, PA 19104, USA}



\begin{abstract}
We construct  new semi-realistic Type IIB flux vacua on $Z_2\times Z_2$
orientifolds with three- and four- Standard Model (SM) families and up to
three units of quantized flux. The open-string sector is comprised of
magnetized D-branes and is T-dual to supersymmetric intersecting D6-brane
constructions. The SM  sector contains magnetized D9-branes with negative
D3-brane charge contribution.  There are large classes of such models and we
present explicit constructions for representative ones. In addition to models
with one and  two units of  quantized flux, we also construct the first three-
and four-family Standard-like models with supersymmetric fluxes, i.e.
comprising three units of  quantized flux. Supergravity fluxes are due to the
self-dual NS-NS and R-R three-form field strength and they fix the toroidal
complex structure moduli and the dilaton. The supersymmetry conditions for the
D-brane sector fix in  some models all three toroidal K\"ahler moduli. We also
provide examples where toroidal K\" ahler moduli are fixed by  strong gauge
dynamics on  the  ``hidden sector'' D7-brane. Most of the models possess Higgs
doublet pairs with Yukawa couplings that can generate masses for quarks and
leptons. The models have (mainly right-) chiral exotics.
\end{abstract} \maketitle

\newpage

\section{Introduction}

One of the  challenging and essential problems in string theory is the
construction of  realistic string vacua, which can stabilize the moduli
fields, generate Standard Model (SM)-like
 gauge structure and  induce a (de Sitter) cosmological constant with  supersymmetry
 breaking.
Such constructions would  provide a bridge between string theory and realistic
particle physics. M-theory provides a framework where, in addition to
perturbative heterotic string vacua, the physical  string vacua could be
probed in the perturbative Type I, Type IIA and Type IIB superstring theory.
In particular, the discovery of D-brane dynamics makes it possible to
construct consistent four-dimensional supersymmetric $N=1$ chiral models with
non-Abelian gauge symmetry on Type II orientifolds, by employing conformal
field theory techniques in the open string sector. The first such
supersymmetric models were based on $Z_2\times Z_2$ orientifolds
\cite{CSUI,CSUII}. [Non-supersymmetric constructions were given in
\cite{LU,IB,IBII,LUII} (see also \cite{Sagnottietal} and for earlier work
\cite{Bachas,BDL}).] Subsequently, a number of   SM-like models, GUT models
and their variations have been constructed in various orbifold backgrounds,
and the associated phenomenology has been discussed. [For a partial list of
non-supersymmetric constructions, see \cite{List}, further supersymmetric
constructions are given in  \cite{ListSUSY}-\cite{CLLL} and further
developments in connection with the  study of effective  couplings and
phenomenological implications, see \cite{PhenSUSY}-\cite{Lust} and references
therein.]

 [Recently important progress has been made  in constructions of
 supersymmetric chiral  solutions of Type II  Gepner models; see \cite{GepnerI,GepnerII} and
references therein. Specifically, the recent impressive results of \cite{GepnerII}
provide large classes of three-family Standard-like Models with no chiral
exotics. Note however, that  these exact conformal field theory models are
located at the special points in  the moduli space where the geometric picture
is lost.  In particular  couplings, such as Yukawa couplings, do not possess
hierarchies associated with the size of the internal spaces, such as in the
case of the toroidal orbifolds with D-branes. In addition, due to the lack of
geometric interpretation, the introduction of supergravity fluxes does not
seem to be  possible. ]

In spite of these successes, the moduli stabilization in open string and
closed string sectors remained an open problem, even though in some cases some
complex structure moduli (in  the Type IIA picture)  and dilaton fields may be
stabilized due to non-perturbative gauge dynamics, associated with the gaugino
condensation in the hidden sector (see, e.g., \cite{CLW}.) Turning on
supergravity fluxes introduces a supergravity potential,  which provides
another  way to stabilize the compactification moduli fields by lifting
continuous moduli space of the string vacua in the effective four-dimensional
theory (see, e.g., \cite{GVW}). However, the introduction of supergravity
fluxes imposes strong constraints on consistent constructions, since such
fluxes modify the global Ramond-Ramond (RR) tadpole cancellation conditions.
Meanwhile, the fluxes will typically generate a back-reaction on the original
geometry of the internal space, thus changing the nature of the internal
space.

On the Type IIA side the supersymmetry conditions of flux compactifications
are less understood. Nevertheless recent work \cite{BCI,BCII} revealed  the
existence of unique flux vacua for massive Type IIA string theory  with SU(3)
structure, whose geometry of the internal six-dimensional space is
nearly-K\"ahler and four-dimensional space is anti-de Sitter (AdS) (for the
discussion on necessary and sufficient conditions of N=1 compactifications of
massive IIA supergravity to AdS(4) with SU(3) structure,  see also
\cite{Lust:2004ig}). One such example is the $ \ {SU(2)^3 \over SU(2)} \simeq
S_3 \times S_3$ coset space that has three supersymmetric three-cycles that
add up to zero in homology \cite{Denefetal,BCII}. Therefore the total  charge
of the D6-branes wrapping such cycles is zero  and no introduction of
orientifold planes on such spaces is needed. Moreover, since the three-cycles
intersect pair-wise,  the massless chiral matters appear at these
intersections. This construction \cite{BCII} therefore provides an explicit
example of supersymmetric flux compactifications  with intersecting D6-branes.
Further progress has  also been made in the construction of $N=1$
supersymmetric Type IIA flux vacua with SU(2)  structures \cite{BCII}, leading
to examples with  the internal space conformally  Calabi-Yau. However,
explicit constructions of  models with intersecting probe D6-branes for  such
flux compactifications is  still awaiting further study.

On the Type IIB side the intersecting D6-brane constructions correspond to models with
magnetized branes with  the role of the  intersecting angles played by the magnetic
fluxes on the branes. The dictionary for  the consistency  and supersymmetry conditions
between the two T-dual constructions is straightforward, see e.g., \cite{CU,BLT}. The
supersymmetric Type IIB flux compactifications are  also better understood; see, e.g.,
\cite{GKP}-\cite{IIB ZN}, \cite{CU,BLT} and references therein. In particular, examples
of supersymmetric fluxes  and the internal space conformally Calabi-Yau are well known.
The prototype example is a self-dual combination of the Neveu-Schwarz-Neveu-Schwarz
(NS-NS) $H_3$ and RR $F_3$ three-forms, corresponding to the primitive (2,1) form on
Calabi-Yau space. Since the back-reaction of such flux configurations is mild, {i.e.},
the internal space remains conformal to Calabi-Yau, these  Type IIB flux
compactifications are especially suitable for adding the probe magnetized D-branes in
this background. However, the quantization conditions on fluxes and the modified
tadpole cancellation conditions constrain the possible D-brane configurations severely.
In Refs. \cite{CU,BLT} the techniques  for consistent chiral flux compactifications on
orbifolds were developed, however, no explicit supersymmetric chiral SM-like models
were obtained.

Most recently, by introducing magnetized D9-branes carrying negative D3-brane
charges in the hidden sector,  in Ref. \cite{MS}  the first example of
three-family SM-like string vacuum with one unit of  quantized flux turned on
was obtained,
 and subsequently, the first four-family SM-like string vacuum with one
unit of fluxes was  constructed in \cite{CL}. These constructions could be
T-dual to the supersymmetric models of intersecting D6-branes on $Z_2\times
Z_2$ orientifold with the $Sp(2)_L\times Sp(2)_R$ or $Sp(2f)_L\times Sp(2f)_R$
gauge symmetry in the electroweak sector, respectively \cite{CIM,CLLL}.
[Without fluxes, the first models of that type were toroidal models with
intersecting D6-branes \cite{CIM} where the RR tadpoles were not explicitly
cancelled. (For the subsequent generalization to tilted tori see
\cite{Kokorelis:2003jr}.)
 The $Z_2\times Z_2$ orientifold construction in
\cite{CLLL}  provided the first model of that type that cancelled RR-tadpoles
by introducing an additional stack of D6-branes with unitary symmetry; in
the T-dual picture those are precisely the magnetized D9-branes with the
negative contribution to the D3-brane charge.] In spite of these successes, we
are confronted by a number of problems:

{(i)} The semi-realistic SM-like string vacua with four-dimensional
 $N=1$ supersymmetric fluxes  have not been constructed, yet.
In view of this drawback, effects of non-supersymmetric fluxes, as a key
mechanism for breaking  supersymmetry,  has been addressed \cite{Soft
BreakingII, Soft BreakingI, Soft BreakingIII}. This analysis  \cite{Soft
BreakingII}  leads to soft supersymmetry breaking masses $M_{soft}\sim
\frac{M_s^2}{M_{Pl}}$ where $M_s$ and $M_{Pl}$ are the string scale and Planck
scale, respectively. This result implies an intermediate string scale or  one
has to introduce an inhomogeneous warp factor in the internal space in order
to stabilize the electroweak scale.

{(ii)} These flux vacua  stabilize  the dilaton and  toroidal complex
structure moduli. However,  the K\"ahler moduli do not enter the flux-induced
superpotential and hence are hard to be completely fixed. So, we are still
typically faced with the vacuum degeneracy problem. [The K\"ahler moduli
fields in Type IIB string theory are T-dual to the complex structure moduli in
Type IIA string theory (intersecting D6-brane scenario). In the T-dual
picture, the latter moduli can often be stabilized by employing
non-perturbative dynamical mechanism, such as the gaugino and matter
condensation in the hidden sector. However, these mechanisms are difficult to
employ  on the Type IIB side due to the additional matter content on the
associated magnetized D-branes.]

{(iii)} For explicit Type IIB orientifolds the  imaginary self-dual fluxes are
quantized in rather large flux units, e.g., for $Z_2\times Z_2$ the orientifolds elementary
flux unit is 64. Therefore the constructions of  semi-realistic flux vacua is very
constrained; only one unit of flux is allowed for known semi-realistic three- \cite{MS}
and four-family \cite{CL} models. Thus the introduction of fluxes is restricted to very
few known semi-realistic examples and is not typical.

Following our previous work~\cite{CL}, in this paper  we systematically study
new constructions of three- and four-family SM-like string vacua with
supergravity fluxes on Type IIB $Z_2\times Z_2$ orientifolds. The major
technical difficulty in constructions of semi-realistic flux  vacua
on Type IIB orientifolds is to ensure the cancellation of the large
positive D3-brane charge contribution to RR tadpoles by the fluxes.  Similar
to the D-brane models without fluxes ~\cite{CLLL}, and the subsequent work
with fluxes \cite{MS,CL}, the important role in tadpole cancellation is
played (in the Type IIB picture) by magnetized D9-branes which carry large
negative D3-brane charges. In the past constructions (\cite{CLLL,MS,CL}) such
D9-branes were introduced as a part of the  ``hidden sector''.

In this paper, we  consider new types of constructions where {\it the magnetized
D9-branes with large negative D3-brane charges are introduced as a part of the SM
sector}. For this new setup, we find that the constructions of SM-like  flux vacua are
much less constrained and obtained a large class of new models. In particular, in
addition to many new models with one unit of quantized flux, {\it we obtain first
three- and four-family models with two units of quantized charge, as well as the first
three- and four-family examples with supersymmetric flux, i.e. three units of quantized
flux}. Such supersymmetric three- and four-family SM-like models have  toroidal
K\"ahler moduli fixed by supersymmetry conditions and the string scale can be close to
the Planck scale. These models have (mainly right-) chiral exotics. However, most of
the models have Higgs doublet pairs with Yukawa couplings to quarks and leptons and
thus  may generate the SM fermion mass hierarchies, explain the CKM quark mixing matrix
and the PMNS neutrino mixing matrix at the tree level, and even give large masses to
some of the bi-fundamental chiral exotics. Finally, with the open-string moduli fixed
by the flux-induced soft masses, we are able to construct the first SM-like string
vacua with strong infrared gauge dynamics on the ``hidden sector'' D7-branes, which
leads to gaugino condensation and provides examples where the third toroidal K\"ahler
modulus is fixed by the strong gauge dynamics \`a la KKLT \cite{KKLT}.

The paper is organized in the following way. In Section II, we systematize the
constructions of supersymmetric string vacua with supergravity fluxes on Type
IIB $Z_2 \times Z_2$ orientifolds. In Section III, we classify the classes of
the SM-like flux vacua  on Type IIB $Z_2 \times Z_2$ orientifolds.
Subsequently,  we discuss in detail explicit constructions of the
representative  models with one, two and three (supersymmetric) units of
quantized  flux, as well as a specific construction with
 gaugino condensation on the ``hidden sector'' D7-branes.
In the Appendix we provide tables of all the explicitly constructed representative
models. We conclude with discussions and open problems in Section IV.

\section{Magnetized D-branes and Type IIB Flux Compactifications
on $T^6 /(Z_2 \times Z_2)$ Orientifolds}

Flux compactifications on simplest  toroidal $T^6$ Type IIB orientifolds, on
which most of the previous work has focused, are unlikely to provide a
framework for constructions of semi-realistic flux vacua. We shall therefore
focus on the simplest  orbifold constructions,  { i.e.} on $T^6 /(Z_2 \times
Z_2)$ orientifolds.

 The internal space $T^{6}$ is chosen to  be factorized as a direct product
of three two-tori,  i.e.  $T^{6} = T^{2} \times T^{2} \times T^{2}$, whose
complex coordinates are $z_i$, $i=1,\; 2,\; 3$ for the $i$-th two-torus,
respectively. The generators $\theta$ and $\omega$ for the orbifold group
$Z_{2} \times Z_{2}$, act on the complex coordinates of $T^6$ as
\begin{eqnarray}
& \theta: & (z_1,z_2,z_3) \to (-z_1,-z_2,z_3)~,~ \nonumber \\
& \omega: & (z_1,z_2,z_3) \to (z_1,-z_2,-z_3)~.~\, \label{orbifold}
\end{eqnarray}
On $Z_2\times Z_2$ orbifold Type IIB string theory contains in the untwisted
sector the four-dimensional $N=2$ supergravity multiplet, the dilaton
hypermultiplet, $h_{11}$ hypermultiplets and $h_{21}$ vector multiplets, which
are all  massless. The orbifold action projects out several components of the
metric of a general $T^6$ geometry and, as a result, we are left with fewer
K\"ahler and complex structure parameters. These are encoded for  the
untwisted  moduli in terms of  the Hodge numbers, as $(h_{11}, h_{21})_{\rm
unt} = (3,3)$. On the other hand, each of the three elements $\theta$,
$\omega$ and $\theta\omega$ has a fixed-point set given by 16 $T^2$'s, and the
corresponding twisted sectors  also contribute to the Hodge numbers of the
orbifold. For a particular choice of discrete torsion, this contribution is
given by $(h_{11}, h_{21})_{\rm tw} = (0,3 \times 16)$. The contributions from
both, the untwisted and twisted sectors hence add up to $(h_{11}, h_{21}) =
(3,51)$.

Orientifold planes are necessary for the introduction of the open-string
sector, and the associated orientifold projection can be denoted by $\Omega
R$, where $\Omega$ is the world-sheet parity projection and $R$ (acting on
Type IIA  as the holomorphic $Z_2$ involution) acts on the complex coordinates
as:
\begin{eqnarray}
R: (z_1,z_2,z_3) \to (-z_1,-z_2,-z_3)~.~
\end{eqnarray}
Thus, the model contains 64 $O3$-planes and 4 $O7_i$-planes, transverse to the $i^{th}$
two-torus. This orientifold action projects the above $N=2$ spectrum to an $N=1$
supergravity multiplet, the dilaton chiral multiplet, and 6 untwisted and 48 twisted
geometrical chiral multiplets.

In order to cancel the  negative RR charge contributions, due to these
O-planes, we need to introduce D(3+2n)-branes which are filling up
four-dimensional
 Minkowski space-time and wrapping
2n-cycles on the compact manifold. We choose the  construction with magnetized
D-branes. [A detailed discussion for  toroidal/orbifold compactifications with
magnetized D-branes is  given, e.g., in  \cite{CU}.] Concretely, for one stack
of $N_a$ D-branes wrapped $m_a^i$ times on the $i^{th}$ two-torus $T^2_i$, we
turn on $n_a^i$ units of magnetic fluxes $F_a$ for the center of mass $U(1)_a$
gauge factor on each $T^2_i$, such that
\begin{eqnarray}
m_a^i \, \frac 1{2\pi}\, \int_{T^2_{\,i}} F_a^i \, = \, n_a^i ~.~\,
\label{monopole}
\end{eqnarray}
Hence, the topological information of this stack of D-branes is encoded in
$N_a$-number of D-branes  and the  co-prime number pairs $(n_a^i,m_a^i)$. The
D9-, D7-, D5- and D3-branes  contain 0, 1, 2 and 3 vanishing $m_a^i$s,
respectively. Introducing for the $i^{th}$ two-torus the even homology classes
$[{\bf 0}_i]$ and $[{\bf T}^2_i]$ for the point and the two-torus,
respectively, the vectors of RR charges of the $a^{th}$ stack of D-branes
 and its image are
\begin{eqnarray}
[{ \Pi}_a]\, =\, \prod_{i=1}^3\, ( n_a^i [{\bf 0}_i] + m_a^i [{\bf T}^2_i] ), & [{
\Pi}_a']\, =\, \prod_{i=1}^3\, ( n_a^i [{\bf 0}_i]- m_a^i [{\bf T}^2_i] )~,~
\label{homology class for D-branes}
\end{eqnarray}
respectively. Similarly, for the O3- and $O7_i$-planes appearing on $T^6/(Z_2\times Z_2)$
orientifold which respectively correspond to $\Omega R$, $\Omega R\omega$, $\Omega
R\theta\omega$ and $\Omega R\theta$ O-planes, we have
\begin{eqnarray}
&\Omega R:&[\Pi_{O3}]= [{\bf 0}_1]\times [{\bf 0}_2]\times [{\bf 0}_3]; \nonumber \\
&\Omega R\omega:&[\Pi_{O7_1}]=-[{\bf 0}_1]\times [{T}^2_2]\times
[{T}^2_3]; \nonumber \\
&\Omega R\theta\omega:&[\Pi_{O7_2}]=-[{T}^2_1]\times [{\bf 0}_2]\times [{T}^2_3];\nonumber\\
&\Omega R\theta:&[\Pi_{O7_3}]=-[{T}^2_1]\times [{T}^2_2]\times [{\bf 0}_3].
\label{homology class for O-planes}
\end{eqnarray}
The ``intersection numbers'', which determine the chiral massless spectrum,
are
\begin{eqnarray}
I_{ab}&=&[\Pi_a]\cdot[\Pi_b]=\prod_{i=1}^3(n_a^im_b^i-n_b^im_a^i)~,~\nonumber \\
I_{ab'}&=&[\Pi_a]\cdot\left[\Pi_{b'}\right]=-\prod_{i=1}^3(n_{a}^im_b^i+n_b^im_a^i)~,~\nonumber \\
I_{aa'}&=&[\Pi_a]\cdot\left[\Pi_{a'}\right]=-8\prod_{i=1}^3(n_a^im_a^i)~,~\nonumber \\
I_{aO}&=&\sum_p[\Pi_a]\cdot[\Pi_{Op}]=8(-m_a^1m_a^2m_a^3
+m_a^1n_a^2n_a^3+n_a^1m_a^2n_a^3+n_a^1n_a^2m_a^3) ~,~\label{intersections}
\end{eqnarray}
where $[\Pi_{Op}]=[\Pi_{O3}]+[\Pi_{O7_1}]+[\Pi_{O7_2}]
+[\Pi_{O7_3}]$ is the sum of O3-plane and $O7_i$-plane
 homology classes. Similar to the discussions in \cite{CSUII},
the physical chiral spectrum  should be invariant under the full orientifold
symmetry group and is tabulated in Table \ref{spectrum}.  Flux vacua  on Type
IIB orientifolds with four-dimensional $N=1$ supersymmetry  are primarily
constrained by the  RR tadpole cancellation conditions
 and conditions for $N=1$ supersymmetry in four dimension,  which  we describe in detail in
 the following subsections.

\begin{table}[t]
\caption{General spectrum  for magnetized D-branes on the  Type IIB
$T^6/(Z_2\times Z_2)$ orientifold. The representations in the table refer to
$U(N_a/2)$, the resulting gauge symmetry due to $Z_2\times Z_2$ orbifold
projection. For supersymmetric constructions, scalars combine with fermions to
form chiral supermultiplets. }
\renewcommand{\arraystretch}{1.25}
\begin{center}
\begin{tabular}{|c|c|}
\hline {\bf Sector} & \phantom{more space inside this box}{\bf Representation}
\phantom{more space inside this box} \\
\hline\hline
$aa$   & $U(N_a/2)$ vector multiplet  \\
       & 3 adjoint chiral multiplets  \\
\hline
$ab+ba$   & $I_{ab}$ $(\fund_a,\antifund_b)$ fermions   \\
\hline
$ab'+b'a$ & $I_{ab'}$ $(\fund_a,\fund_b)$ fermions \\
\hline $aa'+a'a$ &$\frac 12 (I_{aa'} - \frac 12 I_{a,Op})\;\;
\Ysymm\;\;$ fermions \\
          & $\frac 12 (I_{aa'} + \frac 12 I_{a,Op}) \;\;
\Yasymm\;\;$ fermions \\
\hline
\end{tabular}
\end{center}
\label{spectrum}
\end{table}

\subsection{RR Tadpole Cancellation Conditions}

In the Type IIB picture the fluxes,  we consider, are due to the  self-dual
three-from field strength, which contributes to the   D3-brane field strength
equation of motion, and thus modifies the D3 charge conservation (RR-tadpole
cancellation) conditions on the compact orientifold. The RR charges, carried
by magnetized D-branes, are classified by their associated homology classes.
Explicitly, for one stack of $N_a$ D-branes with wrapping numbers
$(n^i_a,m^i_a)$, it carries D3-, D5-, D7- and D9-brane RR charges
\begin{eqnarray}
Q3_a=N_an_a^1n_a^2n_a^3,& (Q5_i)_a=N_am_a^in_a^jn_a^k\, , \nonumber\\
(Q7_i)_a=N_an_a^im_a^jm_a^k,&Q9_a=N_am_a^1m_a^2m_a^3\, ,
\end{eqnarray}
where $i\neq j\neq k$, and a permutation is implied for $(Q5_i)_a$ and $(Q7_i)_a$. So,
the RR tadpole cancellation conditions can be described as
\begin{eqnarray}
\sum_a N_a [\Pi_a]+\sum_a N_a
\left[\Pi_{a'}\right]+\sum_pN_{Op}Q_{Op}[\Pi_{Op}]+N_{\rm flux}=0~,~\,
\end{eqnarray}
where the third term contribution comes from the O3- and $O7_i$-planes, with
$N_{Op}$ and $Q_{Op}$ denoting their numbers and RR charges, respectively. And
$N_{\rm flux}$ is the amount  of the fluxes turned on, and is quantized in
units of the elementary flux.

For a supersymmetric Dp/Dp$'$-brane system on Type IIB $T^6/(Z_2\times Z_2)$
orientifold, only D3- and D7-branes are allowed to be wrapped along the
orientifold planes. [In the following we shall refer to this type of branes as
``filler branes''.] Given that $N_{Op}Q_{Op}=2^{9-p}(-2^{p-4})\equiv -32$ in
Dp-brane units for $Sp$-type O-planes, the RR tadpole cancellation conditions
can be further simplified as
\begin{eqnarray}
-N^{(0)}-\sum_a Q3_a-\frac{1}{2}N_{\rm flux}=-16 ~,~ \nonumber \\
-N^{(i)}+\sum_a (Q7_i)_a=-16, ~i\neq j\neq k  \, , \label{RR tadpole}
\end{eqnarray}
where $N^{(0)}$ and $N^{(i)}$ with $i=$1, 2 and 3 respectively denote the
number of filler branes,  i.e. D-branes which wrap along the O3- and
$O7_i$-planes and only contribute to one of the four kinds of D3- and D7-brane
charges. As for D5- and D9-brane RR tadpoles, their cancellations are
automatic since these D-branes and their $\Omega R$ images carry the same
absolute value of the  corresponding charges, but with opposite sign of
charges.

\subsection{Conditions for Four-Dimensional $N = 1$ Supersymmetry}

Four-dimensional $N=1$ supersymmetric vacua from flux compactification require
that 1/4 supercharges  of the  ten-dimensional (T-dual) Type I theory be
preserved in both open string and close string sectors. We shall discuss both
sectors separately.

For the closed string sector, the specific  Type IIB flux solution on
orientifolds  comprises of self-dual three-form field strength and it has been
discussed, e.g., in  \cite{GKP,KST}. While RR $F_3$ and NSNS $H_3$ three-form
fluxes are turned on, the induced three-form $G_3=F_3-\tau H_3$, with
$\tau=a+i/g_s$ being Type IIB axion-dilaton coupling, contributes to the
D3-brane RR charges
\begin{eqnarray}
N_{\rm flux}\, = \, \frac{1}{(4\pi^2\alpha')^2}\, \int_{X_6}\, H_3\wedge F_3 \,=\,
\frac{1}{(4\pi^2\alpha')^2} \, \frac{i}{2\tau_I}\, \int_{X_6} \, G_3\wedge \bar{G}_3~,~\,
\end{eqnarray}
where $\tau_I$ is the imaginary part of the complex coupling $\tau$. Dirac quantization
conditions of $F_3$ and $H_3$ on $T_6/(Z_2\times Z_2)$ orientifold require that
$N_{\rm flux}$ be a multiple of 64, and the BPS-like  self-duality condition: $*_6G_3=iG_3$
ensures that its contribution to the  RR  charges is positive. Supersymmetric
configuration implies that $G_3$ background field should be a primitive self-dual
(2,1) form. A specific supersymmetric solution which is useful for our purpose is
\cite{CU,KST}
\begin{eqnarray}
G_3\, =\textstyle{\frac{8}{\sqrt{3}}}\, e^{-\pi i/6}\, (\, d{\bar{ z}}_1dz_2dz_3\, + \,
dz_1d{\bar {z}}_2dz_3\, +\, dz_1dz_2d{\bar{ z}}_3\,), \label{G3}
\end{eqnarray}
where the additional factor 4 is due to the $Z_2\times Z_2$ orbifold symmetry. The fluxes
stabilize the complex structure toroidal moduli at values
\begin{eqnarray}
\tau_1=\tau_2=\tau_3=\tau=e^{2\pi i/3},
\end{eqnarray}
leading to the RR tadpole  contribution in Eq. (\ref{RR tadpole})
\begin{eqnarray}
N_{\rm flux}=192.
\end{eqnarray}
This result therefore implies that in order to construct  supersymmetric
SM-like flux vacua, we have to introduce at least the D3 charge conservation
condition, which is thus  hard to achieve.

In the open-string sector, for D-branes with world-volume magnetic field
$F^i=\frac{n^i}{m^i\chi^i}$,  the four-dimensional $N=1$ supersymmetry is ensured  if
and only if  $\sum_i\, \theta_i=0 $ (mod $2\pi$) is satisfied \cite{CU}.  Here the ``
angle'' $\theta_i$  with the range $\{0,\, 2\pi\}$, is determined in terms of the
world-volume magnetic field  as $\tan (\theta_i)\equiv
(F^i)^{-1}=\frac{m^i\chi^i}{n^i}$ and $\chi^i=R^i_1R^i_2$, the area of the $i^{th}$
two-torus $T^2_i$ in $\alpha'$ units, is  the K\"ahler modulus of the $i^{th}$
two-torus $T^2_i$. This supersymmetry condition can be cast in the form: $\sum_i\,
(F^i)^{-1}\, -\, (F^1\, F^2\, F^3)^{-1}=0$, along with $\sum_{i<j}\, (F^i\, F^j)^{-1}\,
- 1\, <\, 0$ for $n_i n_j n_k > 0$ or $\sum_{i<j}\, (F^i\, F^j)^{-1}\, - 1\, >\, 0$ for
$n_i n_j n_k < 0$, which  can be rewritten in the following form:
\begin{eqnarray}
-x_AQ9_a+x_B(Q5_1)_a+x_C(Q5_2)_a+x_D(Q5_3)_a=0, \nonumber\\\nonumber \\
-Q3_a/x_A+(Q7_1)_a/x_B+(Q7_2)_a/x_C+(Q7_3)_a/x_D<0, \label{susyconditions}
\end{eqnarray}
where $x_A=\lambda,\; x_B=\lambda/\chi^2\chi^3,\; x_C=\lambda/\chi^1\chi^3,\;
x_D=\lambda/\chi^1\chi^2$. The positive parameter $\lambda$ has been introduced to put
all the variables $Q9$, $Q7_i$, $Q5_i$, and $Q3_i$ on  equal footing. These
supersymmetry conditions can be easily  cast in the T-dual form of the Type IIA
supersymmetry constraints discussed in \cite{CSUII}.

\section{Constructions of SM-like String Vacua from Type IIB flux compactification}

Similar to the past constructions (see, specifically  those of \cite{CLL}), we
construct the SM-like models as descendants of the Pati-Salam model based on
$SU(4)_C\times SU(2)_L\times SU(2)_R$ gauge symmetry. The hypercharge is
\begin{eqnarray}
Q_Y=Q_{I_{3R}}+{{Q_{B-L}}\over{2}}\, , \label{hypercharge}
\end{eqnarray}
where  the non-anomalous $U(1)_{B-L}$ is obtained from the splitting of the
$U(4)_C$ branes,  i.e. $ U(4) \to U(3)_C\times U(1)_{B-L}$. Similarly, the
anomaly-free $U(1)_{I_{3R}}$ gauge symmetry is from the non-Abelian $U(2)_R$
or $Sp(2)_R$ gauge symmetry. There are three main frameworks to realize the
Pati-Salam gauge sector  in the  Type IIB magnetized D-brane scenario (T-dual
to the  Type IIA intersecting D6-brane one):

 {(i)} The starting
observable sector gauge symmetry is $U(4)_C\times U(2)_L\times U(2)_R$
\cite{CLL}. In this framework, the three  ``anomalous'' gauge symmetries
$U(1)_C$, $U(1)_L$ and $U(1)_R$ can  be treated as global ones, since the
associated gauge bosons obtain   masses  via $B\wedge F$ Chern-Simons
couplings. [Those are effective couplings that arise from  the D-brane
world-volume Chern-Simons couplings  and are responsible for the Abelian gauge
anomaly cancellation via the Green-Schwarz mechanism.] The gauge symmetry
breaking chain is of the form:
\begin{eqnarray}
&& SU(4)\times SU(2)_L \times SU(2)_R\nonumber \\
\rightarrow && SU(3)_C\times SU(2)_L
\times SU(2)_R \times U(1)_{B-L} \nonumber\\
\rightarrow && SU(3)_C\times SU(2)_L\times U(1)_{I_{3R}}\times U(1)_{B-L} \nonumber\\
\rightarrow && SU(3)_C\times SU(2)_L\times U(1)_Y~,~\, \label{breaking chain I}
\end{eqnarray}
where the first and second step can be achieved by splitting the $U(4)_C$
branes and $U(2)_R$ branes in one two-torus direction, and the third step by
giving vacuum expectation values (VEVs) to the scalar components of
right-handed neutrino  chiral superfields  (or a scalar component of an exotic
non-chiral chiral superfields) at the TeV scale. Alternatively, we may skip
the second step and directly break the $SU(2)_R \times U(1)_{B-L}$ gauge
symmetry down to the $U(1)_Y$ by giving VEVs to the scalar components of
right-handed neutrino chiral superfields. Within this  framework one typically
obtains enough SM Higgs doublet pairs {(from the chiral or the non-chiral
massless sector)}  with  Yukawa couplings to quarks and leptons and hence
explain the mass hierarchy and mixings for the fermions in the SM sector at
the tree level. However, it should also be noted that in a few cases  all the
SM Higgs doublet pairs have  the global $U(1)$  quantum numbers that do not
allow for Yukawa couplings  to quarks and leptons; these models therefore
 face serious phenomenological obstacles.

{(ii)} The starting gauge symmetry is one-family $U(4)\times Sp(8)_L\times Sp(8)_R$ or
two-family $U(4)\times Sp(4)_L\times Sp(4)_R$, which can be broken down to the
four-family $U(4)\times U(2)_L\times U(2)_R$ or $U(4)\times SU(2)_L\times SU(2)_R$ by
parallel splitting the D-branes, originally positioned on the O-planes, in three or
two-tori directions, respectively. [Both the string theory and field theory aspects of
the brane splittings in this framework are discussed in detail in \cite{CLLL} for
constructions without fluxes. The flux vacua of that type (with one unit of quantized
flux) was constructed in \cite{CL}.] In the field theory picture (``Higgsing'') the
four-families ($f=4$) are obtained when we decompose the original chiral
supermultiplets $(4,8,1)$ and $({\bar 4},1,8)$, or $(4,4,1)$ and $({\bar 4},1,4)$ into
four copies of $(4,2,1)$ and $({\bar 4},1,2)$ after the gauge symmetry breaking. These
Higgsings, as discussed in \cite{CLLL}, preserve the D- and F-flatness,
 and thus the symmetry breaking can take place at the string scale.
Thus, in these cases the resulting  spectrum is that of four-dimensional $N=1$
supersymmetric four-family Pati-Salam models. The symmetry breaking chains for
these two pictures are given  respectively by
\begin{eqnarray}
&& SU(4)\times Sp(8)_L \times Sp(8)_R\nonumber \\
\rightarrow && SU(4)\times U(2)_L \times U(2)_R \nonumber\\
\rightarrow && SU(3)_C\times U(2)_L \times U(2)_R \times
U(1)_{B-L} \nonumber\\
 \rightarrow && SU(3)_C\times U(2)_L\times U(1)_Y~.~\,
\end{eqnarray}
and
\begin{eqnarray}
&& SU(4)\times Sp(4)_L \times Sp(4)_R\nonumber \\
\rightarrow&& SU(4)\times SU(2)_L \times SU(2)_R \nonumber\\
\rightarrow && SU(3)_C\times SU(2)_L \times SU(2)_R \times
U(1)_{B-L} \nonumber\\
 \rightarrow && SU(3)_C\times SU(2)_L\times U(1)_Y~.~\,
\end{eqnarray}
The first and second step can be achieved by splitting the $Sp$- and
$U(4)$-branes at the string scale, and the third step by giving VEVs to the
scalar components of the right-handed neutrino superfields at the TeV scale.
Note that for  the model with  the original $U(4)\times Sp(8)_L\times Sp(8)_R$
symmetry, the  resulting $U(1)_L$ and $U(1)_R$  are not anomalous since they
are part  of  the non-Abelian $Sp$ symmetries. One expects that at least the
gauge boson of $U(1)_L$ will  obtain  a mass at the electroweak scale, which
is excluded by  experiments. In order to evade this problem, we allow only for
the $Sp(8)_R$ gauge symmetry  in the SM sector, while the $Sp(8)_L$ gauge
symmetry is not. Moreover, we consider the variants of the $U(4)\times
Sp(4)_L\times Sp(4)_R$ model,  i.e. the models with the gauge symmetry
$U(4)\times U(2)_L\times Sp(4)_R$ or $U(4)\times Sp(4)_L\times U(2)_R$. [This
analysis can also be applied to three-family models with gauge symmetry
$U(4)\times Sp(6)_L\times Sp(6)_R$ where the $Sp(6)_L\times Sp(6)_R$ gauge
symmetry can be broken down to the $ Sp(2)_L\times Sp(2)_R$ by the Higgs
mechanism, however this symmetry breaking pattern breaks supersymmetry
\cite{CLLL}, and thus may only be implemented within the framework of
supersymmetry breaking at scale larger than  the electroweak scale.]

{(iii)} The  starting symmetry  is the Pati-Salam-like $U(4)_C\times
Sp(2)_L\times Sp(2)_R$. Without fluxes, the first models of that type were
toroidal orientifolds with intersecting D6-branes \cite{CIM} where the RR
tadpoles were not explicitly cancelled. The $Z_2\times  Z_2$ orientifold
construction in Ref. \cite{CLLL}  cancelled the RR-tadpoles by introducing an
additional stack of branes with unitary symmetry in the hidden sector. In
the T-dual Type IIB picture those are  the magnetized D9-branes with  {\it a
negative ${\rm D3}$ charge}. In  Ref. \cite{MS} these  types of magnetized
D9-branes  were employed to find the MSSM-like model with one unit of
quantized flux turned on.  In this framework, the starting gauge symmetry is
similar to the framework (i); the initial framework  gauge symmetry
$U(4)_C\times Sp(2)_L\times Sp(2)_R$ can be broken down to the SM gauge
symmetry as
\begin{eqnarray}
&& SU(4)\times SU(2)_L \times SU(2)_R \nonumber\\
\rightarrow && SU(3)_C\times SU(2)_L \times SU(2)_R \times
U(1)_{B-L} \nonumber\\
 \rightarrow && SU(3)_C\times SU(2)_L\times U(1)_Y~,~\,
\end{eqnarray}
where the first step can be achieved again, by splitting the $U(4)$ D-branes at string
scale and the second step is achieved by giving VEVs to the scalar components of the
right-handed neutrino chiral superfields at the TeV scale. Note that in the present
case there is  no  $U(1)_L$. However, in the models constructed in Ref. \cite{CIM, MS},
there exists only one  SM  Higgs doublet pair; in this case a generic problem is that
only the third family can obtain the tree-level masses and it is difficult to give
masses to the first two families at the quantum level \cite{CIM}. In addition, the
$SU(2)_R \times U(1)_{B-L}$ gauge symmetry can be broken down to the $U(1)_Y$ only by
giving VEVs to the scalar components of the right-handed neutrino chiral superfields at
the  TeV scale in this kind of models.

The  Pati-Salam-type models with  only  one SM  Higgs pair suffer from serious
phenomenological problems \footnote{We thank Paul Langacker for extensive
discussions on  phenomenological implications of these types of models. See
also \cite{CLLLII}.}. Even though one may be able to generate the most general
Yukawa couplings via radiative corrections, the mass matrix of up-type quarks
is proportional to that of down-type quarks, and the mass matrix of neutrinos
is proportional to that of leptons. (Note the renormalization group equation
running effects on these mass matrices are negligible.) Therefore, on the one
hand, the masses for the quarks, leptons and neutrinos satisfy
\begin{eqnarray}
m_{u}:m_{c}:m_t ~=~ m_d:m_s:m_d~,~ m_{\nu_e} : m_{\nu_{\mu}} :
m_{\nu_{\tau}} ~=~m_e:m_{\mu}:m_{\tau}~.~
\end{eqnarray}
The above fermion mass relations are obviously wrong from the known
experiments. On the other hand, the CKM quark mixing matrix and the PMNS
neutrino mixing matrix are proportional to the identity matrix which implies
that the quark and neutrino mixing angles vanish, again in contradiction with
experiments. Note that these problems for the fermion masses and mixings
cannot be solved by loop corrections  because the $SU(2)_R \times U(1)_{B-L}$
gauge symmetry is broken at TeV scale.  There is a consensus that the minimal
supersymmetric Pati-Salam or Left-Right model should have at least two SM
Higgs doublets \cite{LR}; therefore, the construction (without fluxes) in
\cite{CLLL}, which actually contains two SM Higgs doublets  is the first one
which can realize the embedding of the  supersymmetric Pati-Salam model  with
realistic features in the Type IIA intersecting D-brane scenario, or
equivalently, the T-dual Type IIB magnetized D-brane scenario.

The  presence of fluxes  further complicates  the  constructions of these
types of models: in the Type IIB background, the $G_3$ fluxes  give a large
positive contribution to the D3-brane RR tadpoles, thus making it extremely
hard to satisfy the D3 charge tadpole cancellation conditions by the
magnetized D-brane sectors. In the first model  with one unit of quantized
flux   \cite{MS},
 the large positive  contribution to D3 charges from the flux, is cancelled by
 the  ``hidden sector'' magnetized D9-branes,  carrying negative D3 charges
 (first introduced  for
 vacua without fluxes in \cite{CLLL}).   Four-family models with
 one unit of flux and the starting  SM-sector  gauge symmetry (ii) were
 constructed by introducing a single  stack of magnetized D9-branes with the
 negative D3 charge in \cite{CL}.
 These  very  few specific  semi-realistic
constructions are extremely constrained, thus implying that semi-realistic flux vacua
are hard to come by. In this paper we advance this program in a new direction,  by
introducing  {\it magnetized D9-branes  with negative D3 charges as a part of the
SM-sector} within frameworks (i) and (ii). As a consequence  we obtain a large number
of the  three- and four-family SM-like flux  with as much as three units of flux turned
on. In particular {\it these constructions provide first four-dimensional $N=1$
 supersymmetric SM-like string vacua (i.e. three units of flux) as well as  first examples
 of  semi-realistic SM-like string vacua with two units of
flux} and  many new models with one unit of flux. In addition, we also
obtained SM-like flux models where the ``hidden sector'' D7-brane gauge
dynamics can induce gaugino condensation that can stabilize the third toroidal
K\"ahler modulus (two other toroidal K\"ahler moduli are fixed by
supersymmetry conditions in the D-brane sector); this is an explicit
construction that may realize the KKLT mechanism~\cite{KKLT}.

In the following  subsections we shall present explicit representative  models within
its class.  Within each class there are typically more models and a sizable number of
models within each class has been obtained by running computer code. The representative
model in each class is typically chosen to have a minimal number of chiral (mainly
right) exotics. In the following, we  give a concise description of the representative
models. (Please, refer to the Appendix for tables containing these models and a
detailed explanation of the notation employed.) A concise discussion  of
phenomenological implications of these models will be given in \cite{CLLLII}.

\subsection{Models with Supersymmetric Fluxes}

\begin{table}
[htb] \footnotesize
\renewcommand{\arraystretch}{1.0}
\caption{The chiral spectrum in the open string sector of $Model-T_1-3$.}
\label{spectrum $model-T_1-3$}
\begin{center}
\begin{tabular}{|c||c||c|c|c||c|c|c|}\hline
$Model-T_1-3$ & $U(4)_C\times U(2)_L\times U(2)_R$
& $Q_4$ & $Q_{2L}$ & $Q_{2R}$ & $Q_{em}$ & $B-L$ & Field \\
\hline\hline
$ab$ & $3 \times (4,\overline{2},1)$ & 1 &$-1$& $0$   & $-\frac 13,\; \frac 23,\;-1,\; 0$ & $\frac 13,\;-1$ & $Q_L, L_L$\\
$ab'$ & $1 \times (\overline{4},\overline{2},1)$ & $-1$ &$-1$& $0$   & $\frac 13,\; -\frac 23,\;1,\; 0$ & $-\frac 13,\;1$ & \\
$ac$ & $12 \times (\overline{4},1,2)$ & $-1$ &0& 1    & $\frac 13,\; -\frac 23,\;1,\; 0$ & $-\frac 13,\;1$ & $Q_R, L_R$\\
$ac'$ & $ 10\times (4,1,2)$ & $1$ &0& 1    & $-\frac 13,\; \frac 23,\;-1,\; 0$ & $\frac 13,\;-1$ & \\
$bc$ & $6\times(1,\overline{2}, 2)$ & 0 & $-1$ & 1    & $-1,\;0,\;0,\;1$ & 0 & $H$\\
$bc'$ & $6\times(1,2,2)$ & 0 &1& $1$    & $-1,\;0,\;0,\;1$ & 0 & $H$\\
$b_{\Ysymm}$ & $2\times(1,3,1)$ & 0 & 2 & 0   & $0,\pm 1$ & 0 & \\
$b_{\overline{\Yasymm}}$ & $2\times(1,\overline{1},1)$ & 0 & $-2$ & 0    & 0 & 0 & \\
$c_{\Ysymm}$ & $46\times(1,1,3)$ & 0 &0& $2$   & $0,\pm 1$ & 0 & \\
$c_{\Yasymm}$ & $146\times(1,1,1)$ & 0 &0& 2    & 0 & 0 & \\
\hline
\end{tabular}
\end{center}
\end{table}

In this subsection, we  construct SM-like string vacua with the supersymmetric
flux configuration, i.e. three units of  quantized flux. Again, the key
feature is the introduction of magnetized  D9-branes  with the negative D3
charge, which is a part of the SM-sector.  These are the first  three- and
four-family SM-models within the supersymmetric flux background. The D-brane
configurations of $Model-T_1-3$ and $Model-F_1-3$ are given in Table
\ref{Model-T_1-3} and Table \ref{Model-F_1-3}, respectively. The chiral
spectrum for the three-family model ($Model-T_1-3$) is given in Table
\ref{spectrum $model-T_1-3$}. All three toroidal K\"ahler moduli in these
models are fixed by the supersymmetry conditions for the D-brane sector [Note
that the open string moduli and the K\"ahler moduli can form combined D-flat
directions, corresponding to  the  brane recombination  (see  Ref.
\cite{CSUII}   in the context of specific orientifold compactifications).
However, due to the flux back-reaction it is expected that the open string
moduli could become massive \cite{CU};  in this case the supersymmetry
conditions do stabilize  K\"ahler moduli.] But, the  SM Higgs doublets do not
have Yukawa couplings to quarks and leptons due to the ``wrong'' quantum
numbers under the global $U(1)$ symmetries, and one has to look for new ways
to generate  quark and lepton masses.  We shall  further  discuss  the masses
of the SM families as well as SM chiral exotics in \cite{CLLLII}.

\subsection{Models with Non-supersymmetric Fluxes}

In this subsection, we shall  consider the string vacua with non-supersymmetric fluxes,
 i.e. with two and one units of quantized  fluxes. With fewer units of quantized flux,
there is more freedom in satisfying the tadpole conditions. In the following we only
present some typical three- and four-family models for each phenomenologically
interesting case. Since the gauge symmetry breaking chain for each model can be easily
determined from the analysis at the beginning of this section, we mainly focus on
additional  phenomenological aspects  of these models.

\subsubsection{Two Flux Units Models}

We constructed the first  three- and four-family SM-like string vacua with two
units of quantized  flux  with the representative models: $Model-T_1-2$ (Table
\ref{Model-T_1-2}), $Model-F_1-2$ (Table \ref{Model-F_1-2}), $Model-F_2-2$
(Table \ref{Model-F_2-2}) and $Model-F_3-2$ (Table \ref{Model-F_3-2}).

For the three-family model ($Model-T_1-2$), its   chiral spectrum is  given in Table
\ref{spectrum $model-T_1-2$}. Note that there is one left-chiral exotic
$(\overline{4},\overline{2},1,1)$, which  has Yukawa couplings to the right-chiral ones
and  SM Higgs doublet pairs (Higgs bidoublets), thus these exotics can obtain a mass at
the electroweak scale. In addition, there are
 five pairs of (non-chiral) SM Higgs doublet pairs, arising
in the $bc$ sector when the $b$ and $c$ stacks of D-branes  are coincident on the first
two-torus; these SM Higgs doublet pairs have correct global U(1) quantum numbers,
allowing  for the Yukawa couplings to quarks and leptons
 and thus the  SM fermion masses and mixings can be generated at the tree level.

\begin{table}
[htb] \footnotesize
\renewcommand{\arraystretch}{1.0}
\caption{The chiral spectrum in the open string sector of $Model-T_1-2$.}
\label{spectrum $model-T_1-2$}
\begin{center}
\begin{tabular}{|c||c||c|c|c||c|c|c|}\hline
$Model-T_1-2$ & $U(4)_C\times U(2)_L\times U(2)_R\times Sp(4)$
& $Q_4$ & $Q_{2L}$ & $Q_{2R}$ & $Q_{em}$ & $B-L$ & Field \\
\hline\hline
$ab$ & $3 \times (4,\overline{2},1)$ & 1 &$-1$& $0$   & $-\frac 13,\; \frac 23,\;-1,\; 0$ & $\frac 13,\;-1$ & $Q_L, L_L$\\
$ab'$ & $1 \times (\overline{4},\overline{2},1)$ & $-1$ &$-1$& $0$   & $\frac 13,\; -\frac 23,\;1,\; 0$ & $-\frac 13,\;1$ & \\
$ac$ & $8 \times (\overline{4},1,2)$ & $-1$ &0& 1    & $\frac 13,\; -\frac 23,\;1,\; 0$ & $-\frac 13,\;1$ & $Q_R, L_R$\\
$ac'$ & $ 8\times (4,1,2)$ & $1$ &0& 1    & $-\frac 13,\; \frac 23,\;-1,\; 0$ & $\frac 13,\;-1$ & \\
$bc$ (Non-chiral) & $(1,2,2,1)+(1,\overline{2},\overline{2},1)$ & - &  $\pm 1$  & $\pm 1$ & - & - & $H$\\
$bc'$ & $4\times(1,2,2)$ & 0 &1& $1$    & $-1,\;0,\;0,\;1$ & 0 & \\
$a(D7)_2$ & $1\times(\overline{4},1,1,4)$ &-1& 0 & 0  &  $\frac 16,\;-\frac{1}{2}$ & $\frac 13,\;-1$ & \\
$b(D7)_2$ & $2\times(1,2,1,4)$ &0& 1 & 0  & $\pm \frac 12$ & 0 & \\
$c(D7)_2$ & $6\times(1,1,2,4)$ &0& 0 & 1  &  $\pm \frac{1}{2}$ & 0 & \\
$c_{\Ysymm}$ & $32\times(1,1,3)$ & 0 &0& $2$   & $0,\pm 1$ & 0 & \\
$c_{\Yasymm}$ & $112\times(1,1,1)$ & 0 &0& 2    & 0 & 0 & \\
\hline
\end{tabular}
\end{center}
\end{table}

For the four-family models, there do not exist any left chiral exotics.
Similar to the $Model-T_1-2$, the suitable SM fermion masses and mixings can
be obtained due to the Yukawa couplings of the SM Higgs doublet pairs in
$Model-F_3-2$. However, the SM fermion masses and mixings can not be generated
at the tree level in the $Model-F_1-2$ and $Model-F_2-2$ because the SM Higgs
doublet pairs  have wrong quantum numbers under the $U(1)$ global symmetries
and thus no  Yukawa couplings to quarks and leptons.

\subsubsection{One Flux  Unit Models with $U(2)_{L,R}$ Negative D3 Charge Branes}

 When one unit of flux is turned on, there is a wealth of models and
  these constructions of three- or
four-family SM-like flux  vacua can be classified in the following way:

{(1)} $Model-T_1-1$ (Table \ref{Model-T_1-1}) and $Model-F_1-1$ (Table
\ref{Model-F_1-1}). Except for some symmetric and anti-symmetric
representations, these two models do not contain any bifundamental chiral
exotics in the observable sector. Their  chiral spectra are given in Table
\ref{spectrum $model-T_1-1$} and Table \ref{spectrum $model-F_1-1$},
respectively. In particular, even though for the four-family model
($Model-F_1-1$) there is an anomaly-free $U(1)_R$ gauge symmetry, it is broken
 at the ``right-handed'' scale, when  the $SU(2)_R\times U(1)_{B-L}$ gauge symmetry is broken down to the
$U(1)_Y$ by giving VEVs to the scalar components of right-handed neutrino chiral
superfields. However, in these two models, the SM Higgs doublet pairs  do not have
Yukawa couplings to quarks and leptons, due to the wrong global $U(1)$ quantum numbers.

\begin{table}
[htb] \footnotesize
\renewcommand{\arraystretch}{1.0}
\caption{The chiral spectrum in the open string sector of $Model-T_1-1$.}
\label{spectrum $model-T_1-1$}
\begin{center}
\begin{tabular}{|c||c||c|c||c|c|c|}\hline
$Model-T_1-1$ & $U(4)_C\times Sp(2)_L\times U(2)_R \times Sp(4)$
& $Q_4$ & $Q_{2R}$ & $Q_{em}$ & $B-L$ & Field \\
\hline\hline
$ab$ & $3 \times (4,\overline{2},1,1)$ & 1 & $0$   & $-\frac 13,\; \frac 23,\;-1,\; 0$ & $\frac 13,\;-1$ & $Q_L, L_L$\\
$ac$ & $3 \times (\overline{4},1,2,1)$ & $-1$ & 1    & $\frac 13,\; -\frac 23,\;1,\; 0$ & $-\frac 13,\;1$ & $Q_R, L_R$\\
$bc$ & $8\times(1,\overline{2},2,1)$ & 0 & $1$    & $-1,\;0,\;0,\;1$ & 0 & $H$\\
$c(D3)$ & $1\times(1,1,\overline{2},4)$ & 0 & -1  & $\pm \frac 12$ & 0 & \\
$c_{\Ysymm}$ & $23\times(1,1,3,1)$ & 0 & $2$   & $0,\pm 1$ & 0 & \\
$c_{\Yasymm}$ & $73\times(1,1,1,1)$ & 0 & 2    & 0 & 0 & \\
\hline
\end{tabular}
\end{center}
\end{table}

\begin{table}
[htb] \footnotesize
\renewcommand{\arraystretch}{1.0}
\caption{The chiral spectrum in the open string sector of $Model-F_1-1$.}
\label{spectrum $model-F_1-1$}
\begin{center}
\begin{tabular}{|c||c||c|c||c|c|c|}\hline
$Model-F_1-1$ & $U(4)_C\times U(2)_L\times Sp(8)_R \times Sp(8)$
& $Q_4$ & $Q_{2L}$ & $Q_{em}$ & $B-L$ & Field \\
\hline\hline
$ab$ & $4 \times (4,\overline{2},1,1)$ & 1 & $-1$   & $-\frac 13,\; \frac 23,\;-1,\; 0$ & $\frac 13,\;-1$ & $Q_L, L_L$\\
$ac$ & $1 \times (\overline{4},1,8,1)$ & $-1$ & 0    & $\frac 13,\; -\frac 23,\;1,\; 0$ & $-\frac 13,\;1$ & $Q_R, L_R$\\
$bc$ & $4\times(1,\overline{2},8,1)$ & 0 & $-1$    & $-1,\;0,\;0,\;1$ & 0 & $H$\\
$a(D7)_2$ & $2\times(4,1,1,8)$ & 1 & 0  &  $\frac 16,\;-\frac{1}{2}$ & $\frac 13,\;-1$ & \\
$b(D7)_2$ & $4\times(1,\overline{2},1,8)$ & 0 & -1  & $\pm \frac 12$ & 0 & \\
$a_{\overline{\Ysymm}}$ & $2\times(\overline{10},1,1,1)$ & -2 & 0   & $\frac 13, -1$ & $\frac 23,-2$ & \\
$a_{\Yasymm}$ & $2\times(6,1,1,1)$ & 2 & 0   & $\frac 13, -1$ & $\frac 23,-2$ & \\
$b_{\overline{\Ysymm}}$ & $10\times(1,\overline{3},1,1)$ & 0 & $-2$   & $0,\pm 1$ & 0 & \\
$b_{\overline{\Yasymm}}$ & $54\times(1,\overline{1},1,1)$ & 0 & -2    & 0 & 0 & \\
\hline
\end{tabular}
\end{center}
\end{table}

{(2)} In the $Model-T_2-1$ (Table \ref{Model-T_2-1}), $Model-T_3-1$ (Table
\ref{Model-T_3-1}) and $Model-F_3-1$ (Table \ref{Model-F_3-1}), there are four or five
pairs of non-chiral Higgs bidoublets arising from the $bc$ sector which can couple via
Yukawa couplings to quarks and leptons,  which can generate fermion mass hierarchies
and mixings. Some of the  additional chiral exotics can obtain large masses by coupling
to these non-chiral Higgs bidoublets.

Even though these models employ non-chiral Higgs bidoublets to give masses to
fermions,  we can easily find  models with chiral Higgs bidoublets with
appropriate Yukawa couplings to quarks and leptons,  see, e.g., $Model-F_1-2$
(Table(\ref{Model-F_1-2})).

Another interesting four-family model is $Model-F_4-1$ (Table \ref{Model-F_4-1}); it
does not  contain any  chiral exotics in the observable sector. The chiral Higgs
bidoublets in the $bc$ sector allow for the Yukawa couplings to quarks and leptons and
thus the fermion masses and mixings can be generated at the tree level. Note also that
the $U(4)_C$ symmetry emerges by $Sp(16)$ D7-brane splitting at the string scale and
thus $U(1)_C$ gauge symmetry is non-anomalous and  is broken at the ``right-handed''
scale by the  VEVs of the scalar components of the right-handed neutrino chiral
superfields.

\subsubsection{One Flux Unit Models with $U(4)_{C}$  Negative D3 Charge Branes}

In this construction the $U(4)_C$  is due to the negative D3 charge magnetized
D9-branes. $Model-T_4-1$ (Table \ref{Model-T_4-1}) and $Model-F_5-1$ (Table
\ref{Model-F_5-1}) are such three- and four-family SM-like models. In both
models, the $SU(2)_R$ gauge symmetry is generated by D7-brane splitting, which
yields eight and four copies of right-chiral representations, respectively. In
particular, the four-family model ($Model-F_5-1$), whose chiral spectrum is
given in Table \ref{spectrum $model-F_5-1$}, have several nice
phenomenological features:

{(i)} No additional bifundamental chiral exotics in the observable sector;

{(ii)} The suitable fermion masses and mixings can be generated by tree level Yukawa
couplings to  the SM Higgs doublet pairs;

{(iii)} No additional electroweak scale $U(1)$ symmetry.

\begin{table}
[htb] \footnotesize
\renewcommand{\arraystretch}{1.0}
\caption{The chiral spectrum in the open string sector of $Model-F_5-1$.}
\label{spectrum $model-F_5-1$}
\begin{center}
\begin{tabular}{|c||c||c|c||c|c|c|}\hline
$Model-F_5-1$ & $U(4)_C\times U(2)_L\times Sp(8)_R \times USp(4)$
& $Q_4$ & $Q_{2L}$ & $Q_{em}$ & $B-L$ & Field \\
\hline\hline
$ab$ & $4 \times (4,\overline{2},1,1)$ & 1 & $0$   & $-\frac 13,\; \frac 23,\;-1,\; 0$ & $\frac 13,\;-1$ & $Q_L, L_L$\\
$ac$ & $1 \times (\overline{4},1,8,1)$ & $-1$ & 1    & $\frac 13,\; -\frac 23,\;1,\; 0$ & $-\frac 13,\;1$ & $Q_R, L_R$\\
$bc$ \ (Non-chiral) & $(1,2,8,1)+(1,\overline{2},8,1)$ & - &  $\pm 1$  & - & - & $H$\\
$a(D7)_1$ & $4\times(4,1,1,4)$ & 1 & 0  & $\frac 16,\;-\frac{1}{2}$ & $\frac 13,\;-1$ & \\
$a_{\Ysymm}$ & $2\times(10,1,1,1)$ & 2 & 0   & $\frac 13, -1$ & $\frac 23,-2$ & \\
$a_{\Yasymm}$ & $30\times(6,1,1,1)$ & 2 & 0   & $\frac 13, -1$ & $\frac 23,-2$ & \\
$b_{\Ysymm}$ & $2\times(1,3,1,1)$ & 0 & $2$   & $0,\pm 1$ & 0 & \\
$b_{\overline{\Yasymm}}$ & $2\times(1,\overline{1},1,1)$ & 0 & -2    & 0 & 0 & \\
\hline
\end{tabular}
\end{center}
\end{table}

\subsection{Models with K\"ahler Moduli Stabilized by D7-brane  Gauge Dynamics}

In this subsection,  we consider the possibility of realizing a stabilization
of the toroidal K\"ahler moduli stabilization \` a la  KKLT  mechanism
\cite{KKLT}.  In the original paper, the KKLT vacua are achieved via three
steps:

{(1)} Turning on self-dual three-form fluxes  on Type IIB Calabi-Yau manifold,
the flux-induced superpotential will fix the dilaton and all the complex
structure moduli;

{(2)} Introducing a K\"ahler-moduli dependent non-perturbative superpotential,
 due to D7-brane strong infrared  gauge dynamics  or Euclidean D3-brane
instanton effect. The vacuum is a supersymmetric anti-de Sitter  one with the
K\"ahler moduli fixed;

{(3)} Adding a set of anti-D3-branes to lift the anti-de Sitter  vacuum to a
de Sitter one.

We will only focus on the second step,  i.e. the generation of the
non-perturbative superpotential, since it may help us stabilize all the
toroidal  moduli fields.

In our framework,  the Type IIB  SM-like flux vacua typically require a stack
of (filler) D3-branes, sitting on the O3-planes, in order to cancel the D3
charge tadpoles due to  redundant negative D3 charges introduced by the
magnetized D9-branes. This stack of D3-branes has  typically a negative beta
function; it thus possesses  a non-perturbative gauge dynamics that results in
a non-perturbative superpotential, due to gaugino condensation. However, this
superpotential depends only on the dilaton-axion field,  and is independent
K\"ahler-moduli. Thus, in order to generate the non-perturbative
superpotential that depends on the toroidal K\"ahler moduli, the strong gauge
dynamics has to arise due to D7-branes \cite{KKLT}.

Recently, it was suggested \cite{Soft BreakingII} that the flux-induced soft
mass terms may help decouple the open-string moduli on D7-branes, leaving an
infrared gauge theory with strong dynamics on the world-volume of this stack
of D7-branes  (in the hidden sector). However, in the concrete constructions
of SM-like flux  vacua  a stack of hidden sector D7-branes typically do not
have a negative beta function; this is due to the fact   that  the magnetized
D9-branes with  negative D3 charge have large ``intersecting numbers''  with
the D7-branes and thus a large number of chiral matter, charged under D7-brane
gauge symmetry.  Additional chiral matter typically drastically modifies the
infrared gauge dynamics.  [Light chiral matter can influence the
(supersymmetric) gauge dynamics in two key   aspects (see, e.g.,
\cite{SuperQCD} and references therein): (i) the chiral matter contributes to
the beta function and thus affect the infrared dynamics  (phase structure);
(ii) it may induce matter condensation and contribute to the non-perturbative
superpotential.]

In this paper, we present the first SM-like flux vacua with strong gauge dynamics,
resulting in  gaugino condensation, on D7-branes of the hidden sector: $Model-T_5-1$
(Table \ref{Model-T_5-1}) and $Model-F_6-1$ (Table \ref{Model-F_6-1}). These models
have three- and four-family fermions (and additional  chiral exotics), respectively. In
these models the two (out of three)  toroidal K\"ahler moduli have been fixed by
supersymmetry conditions. For the four-family model ($Model-F_6-1$), its hidden sector
is composed of two stacks of D7-branes denoted by $(D7)_1$ and $(D7)_2$, respectively.
Both of them carry $Sp(4)$ gauge symmetry and the associated beta functions are -3(0)
and -5(-2), respectively. Here the beta functions in the brackets include the one-loop
contribution from the open-string moduli. These open-string moduli are expected to
become massive due to the flux back-reaction (see, e.g.,\cite{CU}), and  in this case
the strong  gauge dynamics can  in principle induce the non-perturbative
superpotential. Note, however, that gauge dynamics of the $(D7)_1$-branes results in
the superconformal or Coulomb phase regime and the non-perturbative superpotential
cannot be dynamically generated. As a result, only $(D7)_2$-branes can generate a
Veneziano-Yankielowicz-type  superpotential,  induced by gaugino and matter
condensations.
 A similar analysis can be applied to the three-family model
($Model-T_5-1$):  in the case that  open string moduli are decoupled (the beta
function is -9(0)), the Veneziano-Yankielowicz superpotential is   generated
due to the gaugino and matter condensates on the D7-brane.

\section{Discussions and Conclusions}

In this paper we have advanced a program  for explicit constructions of
supersymmetric  SM-like  string vacua with supergravity fluxes turned on. In
particular we obtained  large classes of  such  Type IIB models on $Z_2\times
Z_2$ orientifolds with one, two, and three units of quantized flux turned on.
These models provide an important stepping stone toward broader classes of
realistic constructions that not only contain the three- (or four-) family SM
sector, but also stabilize some of the  moduli (typically all toroidal moduli
can be fixed).

Before our work, techniques for constructions of the chiral D-brane sectors
flux vacua on Type IIB orientifolds were developed in \cite{CU,BLT}, however,
the semi-realistic constructions remained elusive until recently. The
technical reason for such difficulties are large positive  three-form flux
contributions to the D3 charge in the internal space (D3-branes RR-tadpole),
which makes the D3 charge conservation constraint hard to satisfy. In the
first  examples of SM flux vacua on $Z_2\times Z_2$ orientifolds  this
constraint has been satisfied  \cite{MS,CL} by the introduction of  the hidden
sector magnetized D9-branes which carry negative D3 charges. [This type of
D-branes were first introduced in the Type IIA context with intersecting
D6-branes without fluxes in \cite{CLLL}.] Within this framework  one
three-family  \cite{MS} and one four-family \cite{CL} SM-like models with one
unit of quantized flux were obtained.  However, one remains to be confronted
by
 serious problems:

{(i)} There remains the  outstanding problem of constructing semi-realistic
SM-like string vacua with  supersymmetric fluxes, that for $Z_2\times Z_2$
orientifolds correspond to three units of quantized  flux.  The focus
therefore shifted to the study of non-supersymmetric flux effects  as the key
to break supersymmetry, and a detailed study of  soft supersymmetry breaking
masses due to fluxes \cite{Soft BreakingII, Soft BreakingI}. The flux-induced
supersymmetry breaking, however,  leads typically to soft masses
$M_{soft}\sim\frac{M_s^2}{M_{Pl}}$, which implies an intermediate string scale
or inhomogeneous warp factor in the internal space in order to stabilize the
electroweak scale;

{(ii)} In spite of the successes in stabilizing the dilaton and toroidal
complex structure moduli, the toroidal  K\"ahler moduli are not completely
fixed, thus one remains faced with the large
 vacuum degeneracy problem;

{(iii)} The fact that  (supersymmetric) supergravity fluxes  are abundant,
makes it is imperative  to extend the constructions to semi-realistic models
with more than one unit of quantized  flux.

 In this paper, we have made  important progress in addressing the above issues.
The key ingredient in the new SM-like flux constructions is the introduction of the
negative D3 charge magnetized D9-brane as a part of the SM sector. These constructions
turned out to be less constraining and resulted in three- and four-family SM-like
string vacua with up to three units of quantized flux, thus leading to the first fully
supersymmetric SM-like  flux vacua. In addition to  toroidal complex structure moduli
and the dilaton being fixed by the flux, the supersymmetry conditions in the D-brane
sector typically fix all the toroidal K\"ahler moduli,  and the string scale is close
to  the Planck scale. We  also constructed the first three- and four-family SM-like
models with two units of the  quantized flux and many  new models with one unit of the
quantized flux. Typically the  representative models  have (mainly right-) chiral
exotics in the SM sector; however,  we have also presented a few three- and four-family
models with no SM-sector chiral exotics. (Note, the models do have additional tensor
fields,  i.e.   chiral superfields in the symmetric and/or anti-symmetric
representation of the unitary gauge symmetry, typically associated with the negative D3
charge magnetized D9-brane.)  We have also been able to construct SM-like flux vacua
with strong infrared gauge dynamics on the hidden sector D7-branes, which leads to
non-perturbative superpotential that can fix a remaining toroidal K\"ahler modulus.

 The SM Higgs doublet pairs
 appear at the intersections of the $SU(2)_L$ and $SU(2)_R$.
  For most models,  the SM fermion   masses and mixings can be generated
due to the tree-level Yukawa couplings of such  SM Higgs doublet pairs to quarks and
leptons.  However, in a few cases, most notably for the three-family SM-like model with
the supersymmetric flux, all such Higgs fields have the wrong global $U(1)$ quantum
numbers that prevent them from coupling
 to quarks and leptons via Yukawa couplings;  these models therefore face serious
 phenomenological difficulties. Note also that typically in these models some of the
SM chiral exotics can obtain masses at least at the electroweak scale, due to the
Yukawa couplings of the SM Higgs doublet pairs to such exotics.

In spite of a number of advances made  with these new constructions, there are
open problems  which deserve further study. In particular, further discussion
of the specific Yukawa couplings to quarks, leptons and chiral exotics, as
well as the subsequent further implications for the masses and mixings of the
SM chiral fermions are needed \cite{CLLLII}. For SM-like models with
supersymmetric fluxes, one has to address the supersymmetry breaking
mechanism. In principle the supersymmetry breaking could be due to the hidden
D-brane strong gauge dynamics; however, the present models do not possess such
a sector. Complete stabilization of all the moduli (and not only the toroidal
closed sector one) remains an open problem.  We postpone these  issues for
future research.


\section*{Acknowledgments}

We would like to thank Paul  Langacker for many discussions,  encouragement, and
collaboration on topics related to this paper. The research was supported in part by
the National Science Foundation
under Grant No.~INT02-03585 (MC) and PHY-0070928 (T. Li), by the
Department of Energy Grant DOE-EY-76-02-3071 (MC, T. Liu), and by the Fay R. and Eugene L.
Langberg Chair  (MC).

\newpage


\begin{center}
\Large{\bf Appendix: D-brane Configurations and Intersection Numbers
for SM-like Flux Vacua}
\end{center}

In this Appendix, we tabulate D-brane configurations and intersection numbers
for the representative  three- and four-family models within our new setup. In
the first column of each table, $a$, $b$ and $c$ denote the $U(4)$ ($Sp(16)$),
$U(2)_L$ ($Sp(2)_L$ or $Sp(4)_L$), and $U(2)_R$ ($Sp(4)_R$ or $Sp(8)_R$)
stacks of branes, respectively. $D3$, $(D7)_1$, $(D7)_2$, and $(D7)_3$
represent the filler branes along respective $\Omega R$, $\Omega R\omega$,
$\Omega R\theta\omega$ and $\Omega R\theta$ orientifold planes, resulting in
$Sp(N)$ gauge groups. $N$, in the second column, corresponds to the number of
filler D-branes in each stack.   The third column  depicts the wrapping
numbers of the various D-branes. The intersection numbers between the various
D-brane stacks are given in the remaining right columns where $b'$ and $c'$
are respectively the $\Omega R$ images of $b$ and $c$.In addition, the number
of symmetric and antisymmetric chiral superfield representations for specific
D-brane configurations is given. For convenience, we also tabulate the
relations among the three toroidal K\"ahler moduli parameters $\chi_i$,
imposed by the supersymmetry conditions. The model labels
``$Model-(T,~F)_i-n$'' appearing in the tables denote the ``$Model-(three, ~
four ~ family)_i-(n ~ units ~ of ~fluxes)$ ''.  [Since all the models have an
even number of chiral supermultiplets  in the fundamental representation of
the $Sp(N)$ gauge groups, these models are automatically free of discrete
global gauge anomalies \cite{Witten}.]  Finally, we emphasize that in this
paper we do not fix the convention between the chirality and the sign of the
intersection number. Instead, we consider the $SU(2)$ D-branes that  carry the
more realistic chiral  spectrum (typically only three- or four- families) as
the $SU(2)_L$ D-branes.  These representative models therefore possess
(mainly) right- chiral exotics.

\begin{table*}
[htb] \footnotesize
\renewcommand{\arraystretch}{1.0}
\caption{D-brane configurations and intersection numbers for $Model-T_1-3$.}
\label{Model-T_1-3}
\begin{center}
\begin{tabular}{|c||c|c||c|c|c|c|c|c||c|}
\hline
    $Model-T_1-3$ & \multicolumn{9}{|c|}{$[U(4)_C\times U(2)_L\times U(2)_R]_{Observable}$}\\
\hline \hline \rm{j} & $N$ & $(n^1,m^1)(n^2,m^2)
(n^3,m^3)$ & $n_{\Ysymm}$& $n_{\Yasymm}$ & $b$ & $b'$ & $c$ & $c'$ &  K\"ahler moduli    \\
\hline \hline
    $a$&  8& $(1,0)(1,1)(1,-1)$ & 0 & 0  &  -3& 1& 12 & -10 & $\chi_3=\chi_2=2\chi_1$ \\
    $b$&  4& $(1,1)(2,-1)(1,0)$ & -2 & 2  &  - & - & 6 & -6&   $\chi_3=2\sqrt{10}$ \\
    $c$&  4& $(-2,-1)(4,1)(3,1)$ & -46 & -146  &  - & - & - &-&    \\
\hline
\end{tabular}
\end{center}
\end{table*}

\begin{table}
[htb] \footnotesize
\renewcommand{\arraystretch}{1.0}
\caption{D-brane configurations and intersection numbers for $Model-T_1-2$.}
\label{Model-T_1-2}
\begin{center}
\begin{tabular}{|c||c|c||c|c|c|c|c|c|}
\hline
    \multicolumn{2}{|c|}{$Model-T_1-2$} & \multicolumn{7}{|c|}
{$[U(4)_C\times U(2)_L\times U(2)_R]_{Observable} \times [Sp(4)]_{Hidden}$}\\
\hline \hline \rm{j} & $N$ & $(n^1,m^1)(n^2,m^2)
(n^3,m^3)$ & $n_{\Ysymm}$& $n_{\Yasymm}$ & $b$  & $b'$ & $c$ & $c'$   \\
\hline \hline
    $a$&  8& $(1,0)(1,1)(1,-1)$ & 0 & 0   & -3 & 1&  8 & -8\\
    $b$& 4& $(2,1)(2,-1)(1,0)$ & 0 & 0  &  - & - &  0 & -4   \\
    $c$&   4& $(-2,-1)(3,1)(3,1)$ & -32 & -112  &  - & - &- & -   \\
\hline \hline
    $(D7)_2$& $4$ & $(0,1)(1,0)(0,-1)$ & \multicolumn{6}{c|} {$\chi_3=\chi_2=\chi_1=\sqrt{21}$} \\
\hline
\end{tabular}
\end{center}
\end{table}

\begin{table}
[htb] \footnotesize
\renewcommand{\arraystretch}{1.0}
\caption{D-brane configurations and intersection numbers for $Model-T_1-1$.}
\label{Model-T_1-1}
\begin{center}
\begin{tabular}{|c||c|c||c|c|c|c|c|}
\hline
    \multicolumn{2}{|c|}{$Model-T_1-1$} & \multicolumn{6}{|c|}{$[U(4)_C\times Sp(2)_L\times U(2)_R]_{Observable} \times [Sp(4)]_{Hidden}$}\\
\hline \hline \rm{j} & $N$ & $(n^1,m^1)(n^2,m^2)
(n^3,\tilde{m}^3)$ & $n_{\Ysymm}$& $n_{\Yasymm}$ & $b$  & $c$ & $c'$    \\
\hline \hline
    $a$&  8& $(1,0)(3,1)(3,-1/2)$ & 0 & 0   & -3 &  3 & 0   \\
    $b$&  2& $(0,1)(0,-1)(2,0)$ & 0 & 0  &  - & 8 & -   \\
    $c$& 4& $(-2,-1)(4,1)(3,1/2)$ & -23 & -73  &   - & - & -   \\
\hline \hline
    $D3$& 4& $(1,0)(1,0)(2,0)$ & \multicolumn{5}{c|}
{$\chi_2=\chi_3$, $\frac{12}{\chi_2^2}+\frac{14}{\chi_1\chi_2}=1$} \\
\hline
\end{tabular}
\end{center}
\end{table}

\begin{table}
[htb] \footnotesize
\renewcommand{\arraystretch}{1.0}
\caption{D-brane configurations and intersection numbers for $Model-T_2-1$.}
\label{Model-T_2-1}
\begin{center}
\begin{tabular}{|c||c|c||c|c|c|c|c|c|}
\hline
    \multicolumn{2}{|c|}{$Model-T_2-1$} & \multicolumn{7}{|c|}{$[U(4)_C\times U(2)_L\times U(2)_R]_{Observable} \times [Sp(4)\times Sp(2)]_{Hidden}$}\\
\hline \hline \rm{j} & $N$ & $(n^1,m^1)(n^2,m^2)
(n^3,\tilde{m}^3)$ & $n_{\Ysymm}$& $n_{\Yasymm}$ & $b$  & $b'$ & $c$ & $c'$    \\
\hline \hline
    $a$&  8& $(1,0)(1,1)(1,-1/2)$ & 0 & 0   & 3 & -2  &  -4 & 4\\
    $b$&  4& $(2,-1)(1,0)(5,1/2)$ & 3 & -3  &  - & - & 0 & -4  \\
    $c$&  4& $(-2,1)(3,-1)(3,-1/2)$ & 16 & 56  &  - & - & - & -  \\
\hline \hline
    $D3$& 4& $(1,0)(1,0)(2,0)$ & \multicolumn{6}{c|} {$\chi_3=\chi_2=\frac{5}{2}\chi_1=\sqrt{39}$} \\
    $(D7)_2$& 2& $(0,1)(0,-1)(2,0)$ & \multicolumn{6}{c|} {} \\
\hline
\end{tabular}
\end{center}
\end{table}

\begin{table}
[htb] \footnotesize
\renewcommand{\arraystretch}{1.0}
\caption{D-brane configurations and intersection numbers for $Model-T_3-1$.}
\label{Model-T_3-1}
\begin{center}
\begin{tabular}{|c||c|c||c|c|c|c|c|c|}
\hline
    \multicolumn{2}{|c|}{$Model-T_3-1$} & \multicolumn{7}{|c|}{$[U(4)_C\times U(2)_L\times U(2)_R]_{Observable} \times [Sp(8)]_{Hidden}$}\\
\hline \hline \rm{j} & $N$ & $(n^1,m^1)(n^2,m^2)
(n^3,m^3)$ & $n_{\Ysymm}$& $n_{\Yasymm}$ & $b$  & $b'$ & $c$ & $c'$    \\
\hline \hline
    $a$&  8& $(1,0)(1,1)(1,-1)$ & 0 & 0   & 3 & 1  &  -3 & 3\\
    $b$&  4& $(2,-1)(1,0)(1,2)$ & -6 & 6  &  - & - & 0 & 12  \\
    $c$&  4& $(-2,1)(2,-1)(2,-1)$ & 10 & 54  &  - & - & - & -  \\
\hline \hline
    $(D7)_3$& 8 & $(0,1)(0,-1)(1,0)$ & \multicolumn{6}{c|}
{$\chi_3=\chi_2=\frac{1}{4}\chi_1=\sqrt{6}$} \\
\hline
\end{tabular}
\end{center}
\end{table}

\begin{table*}
[htb] \footnotesize
\renewcommand{\arraystretch}{1.0}
\caption{D-brane configurations and intersection numbers for $Model-T_4-1$.}
\label{Model-T_4-1}
\begin{center}
\begin{tabular}{|c||c|c||c|c|c|c|c|}
\hline
    \multicolumn{2}{|c|}{$Model-T_4-1$} & \multicolumn{6}{|c|}{$[U(4)_C\times U(2)_L\times Sp(4)_R]_{Observable} \times [Sp(4)]_{Hidden}$}\\
\hline \hline \rm{j} & $N$ & $(n^1,m^1)(n^2,m^2)
(n^3,m^3)$ & $n_{\Ysymm}$& $n_{\Yasymm}$ & $b$  & $b'$ & $c$    \\
\hline \hline
    $a$&  8& $(-1,-1)(2,1)(2,1)$ & -2 & -30  &  3 & -5 & -4 \\
    $b$&  4& $(1,0)(3,1)(1,-1)$ & 4 & -4  &  - & - & 0  \\
    $c$&  4& $(1,0)(0,1)(0,-1)$ & 0 & 0  &  - & - & -  \\
\hline \hline
    $D3$&  4 & $(1,0)(1,0)(1,0)$ & \multicolumn{5}{c|} {$3\chi_3=\chi_2$, $\frac{12}{\chi_2^2}+\frac{8}{\chi_1\chi_2}=1$} \\
\hline
\end{tabular}
\end{center}
\end{table*}

\begin{table*}
[htb] \footnotesize
\renewcommand{\arraystretch}{1.0}
\caption{D-brane configurations and intersection numbers for $Model-T_5-1$, here
$\beta^g$ are beta functions for the associated gauge symmetries in the hidden sector.}
\label{Model-T_5-1}
\begin{center}
\begin{tabular}{|c||c|c||c|c|c|c|c|}
\hline
    \multicolumn{2}{|c|}{$Model-T_5-1$}&\multicolumn{6}{|c|}{$[U(4)_C\times U(2)_L\times Sp(8)_R]_{Observable} \times [Sp(8)\times Sp(8)]_{Hidden}$}\\
\hline \hline \rm{j} & $N$ & $(n^1,m^1)(n^2,m^2)
(n^3,m^3)$ & $n_{\Ysymm}$& $n_{\Yasymm}$   & $b$ & $b'$  & $c$ \\
\hline \hline
    $a$&  8& $(1,0)(1,1)(1,-1)$ & 0 & 0   &  $3$ & $-3$ &$-1$ \\
    $b$&  4& $(-2,-1)(2,1)(2,1)$ & $-10$ & $-54$  &  -& - & $-4$  \\
    $c$&  8& $(0,1)(0,-1)(1,0)$ & 0 & 0  &  -  & - & -  \\
\hline \hline
    $D3$&  8 & $(1,0)(1,0)(1,0)$ & \multicolumn{5}{c|} {$\chi_2=\chi_3$, $\frac{4}{\chi_2^2}+\frac{8}{\chi_1\chi_2}=1$} \\
    $(D7)_2$&  8 & $(0,1)(1,0)(0,-1)$ & \multicolumn{5}{c|} {$\beta^g_{D3}=-14(-5)$, $\beta^g_{(D7)_2}=-9(0)$} \\
\hline
\end{tabular}
\end{center}
\end{table*}

\begin{table*}
[htb] \footnotesize
\renewcommand{\arraystretch}{1.0}
\caption{D-brane configurations and intersection numbers for $Model-F_1-3$.}
\label{Model-F_1-3}
\begin{center}
\begin{tabular}{|c||c|c||c|c|c|c|c||c|}
\hline
    $Model-F_1-3$ & \multicolumn{8}{|c|}{$[U(4)_C\times Sp(4)_L\times U(2)_R]_{Observable}$}\\
\hline \hline \rm{j} & $N$ & $(n^1,m^1)(n^2,m^2)
(n^3,m^3)$ & $n_{\Ysymm}$& $n_{\Yasymm}$ & $b$ & $c$ & $c'$ & K\"ahler moduli    \\
\hline \hline
    $a$&  8& $(1,0)(2,1)(1,-1)$ & 2 & -2  & -2& 8 & -12   & $\chi_2=2\chi_3$ \\
    $b$& 4& $(0,1)(0,-1)(1,0)$ & 0 & 0  &  - &  8 & -& $\frac{24}{\chi_2^2}+\frac{20}{\chi_1\chi_2}=1$   \\
    $c$& 4& $(-2,-1)(4,1)(3,1)$ & -46 & -146  &  - &  - &-&    \\
\hline
\end{tabular}
\end{center}
\end{table*}

\begin{table*}
[htb] \footnotesize
\renewcommand{\arraystretch}{1.0}
\caption{D-brane configurations and intersection numbers for $Model-F_1-2$.}
\label{Model-F_1-2}
\begin{center}
\begin{tabular}{|c||c|c||c|c|c|c|c|c|}
\hline
    \multicolumn{2}{|c|}{$Model-F_1-2$} & \multicolumn{7}{|c|}{$[U(4)_C\times U(2)_L\times U(2)_R]_{Observable} \times [Sp(4)]_{Hidden}$}\\
\hline \hline \rm{j} & $N$ & $(n^1,m^1)(n^2,m^2)
(n^3,m^3)$ & $n_{\Ysymm}$& $n_{\Yasymm}$ & $b$  & $b'$ & $c$ & $c'$    \\
\hline \hline
    $a$&  8& $(1,0)(2,1)(1,-1)$ & 2 & -2   & -4 & 0   &  4 & -10\\
    $b$&  4& $(1,1)(2,-1)(1,0)$ & -2 & 2  &  - & - & 5 & -3  \\
    $c$&  4& $(-2,-1)(3,1)(3,1)$ & -32 & -112  &  - & - & - & -  \\
\hline \hline
    $(D7)_2$& 4 & $(0,1)(1,0)(0,-1)$ & \multicolumn{6}{c|} {$\chi_2=2\chi_3=2\chi_1=3\sqrt{6}$} \\
\hline
\end{tabular}
\end{center}
\end{table*}

\begin{table*}
[htb] \footnotesize
\renewcommand{\arraystretch}{1.0}
\caption{D-brane configurations and intersection numbers for $Model-F_2-2$.}
\label{Model-F_2-2}
\begin{center}
\begin{tabular}{|c||c|c||c|c|c|c|c|}
\hline
    \multicolumn{2}{|c|}{$Model-F_2-2$} & \multicolumn{6}{|c|}{$[U(4)_C\times Sp(4)_L\times U(2)_R]_{Observable} \times [Sp(8)\times Sp(4)]_{Hidden}$}\\
\hline \hline \rm{j} & $N$ & $(n^1,m^1)(n^2,m^2)
(n^3,m^3)$ & $n_{\Ysymm}$& $n_{\Yasymm}$ & $b$  & $c$ & $c'$    \\
\hline \hline
    $a$&  8& $(1,0)(2,1)(1,-1)$ & 2 & -2  & -2 &  4 & -10  \\
    $b$&   4& $(0,1)(0,-1)(1,0)$ & 0 & 0  &  - & 6 & -  \\
    $c$& 4& $(-2,-1)(3,1)(3,1)$ & -32 & -112  &  - & - & -  \\
\hline \hline
    $D3$& 8& $(1,0)(1,0)(1,0)$ & \multicolumn{5}{c|} {$2\chi_3=\chi_2$} \\
    $(D7)_2$& 4 & $(0,1)(1,0)(0,-1)$  & \multicolumn{5}{c|} {$\frac{18}{\chi_2^2}+\frac{18}{\chi_1\chi_2}=1$} \\
\hline
\end{tabular}
\end{center}
\end{table*}

\begin{table}
[htb] \footnotesize
\renewcommand{\arraystretch}{1.0}
\caption{D-brane configurations and intersection numbers for $Model-F_3-2$.}
\label{Model-F_3-2}
\begin{center}
\begin{tabular}{|c||c|c||c|c|c|c|c|c|}
\hline
    \multicolumn{2}{|c|}{$Model-F_3-2$} & \multicolumn{7}{|c|}{$[U(4)_C\times U(2)_L\times U(2)_R]_{Observable} \times [Sp(4)]_{Hidden}$}\\
\hline \hline \rm{j} & $N$ & $(n^1,m^1)(n^2,m^2)
(n^3,m^3)$ & $n_{\Ysymm}$& $n_{\Yasymm}$ & $b$  & $b'$ & $c$ & $c'$    \\
\hline \hline
    $a$&  8& $(1,0)(2,1)(1,-1)$ & 2 & -2   & -3 & -1 &  4 & -10  \\
    $b$& 4& $(2,1)(1,-1)(1,0)$ & 2 & -2  &  - & -  & 0 & -8  \\
    $c$&  4& $(-2,-1)(3,1)(3,1)$ & -32 & -112  &  - & - & - & -  \\
\hline \hline
    $(D7)_2$& 4& $(0,1)(1,0)(0,-1)$ & \multicolumn{6}{c|} {$\chi_2=2\chi_3=\frac{1}{2}\chi_1=3\sqrt{3}$} \\
\hline
\end{tabular}
\end{center}
\end{table}

\begin{table*}
[htb] \footnotesize
\renewcommand{\arraystretch}{1.0}
\caption{D-brane configurations and intersection numbers for $Model-F_1-1$.}
\label{Model-F_1-1}
\begin{center}
\begin{tabular}{|c||c|c||c|c|c|c|c|}
\hline
    \multicolumn{2}{|c|}{$Model-F_1-1$} & \multicolumn{6}{|c|}{$[U(4)_C\times U(2)_L\times Sp(8)_R]_{Observable} \times [Sp(8)]_{Hidden}$}\\
\hline \hline \rm{j} & $N$ & $(n^1,m^1)(n^2,m^2)
(n^3,m^3)$ & $n_{\Ysymm}$& $n_{\Yasymm}$ & $b$  & $b'$ & $c$   \\
\hline \hline
    $a$&  8& $(1,0)(1,1)(2,-1)$ & -2 & 2  &  4 & 0 & -1  \\
    $b$&  4& $(-2,-1)(2,1)(2,1)$ & -10 & -54  &  - & - & -4  \\
    $c$&  8& $(0,1)(0,-1)(1,0)$ & 0 & 0  &  - & - & -  \\
\hline \hline
    $(D7)_2$& 8 & $(0,1)(1,0)(0,-1)$ & \multicolumn{5}{c|} {$\chi_3=2\chi_2$, $\frac{2}{\chi_2^2}+\frac{6}{\chi_1\chi_2}=1$} \\
\hline
\end{tabular}
\end{center}
\end{table*}

\begin{table}
[htb] \footnotesize
\renewcommand{\arraystretch}{1.0}
\caption{D-brane configurations and intersection numbers for $Model-F_2-1$.}
\label{Model-F_2-1}
\begin{center}
\begin{tabular}{|c||c|c||c|c|c|c|c|c|}
\hline
    \multicolumn{2}{|c|}{$Model-F_2-1$} & \multicolumn{7}{|c|}{$[U(4)_C\times U(2)_L\times U(2)_R]_{Observable} \times [Sp(16)\times Sp(4)]_{Hidden}$}\\
\hline \hline \rm{j} & $N$ & $(n^1,m^1)(n^2,m^2)
(n^3,m^3)$ & $n_{\Ysymm}$& $n_{\Yasymm}$ & $b$  & $b'$ & $c$ & $c'$    \\
\hline \hline
    $a$&  8& $(1,0)(1,1)(2,-1)$ & -2 & 2   & 4 & 0 &  -4 & 10 \\
    $b$&  4& $(3,-1)(1,0)(2,1)$ & -2 & 2  &  - & - & 5 & 5  \\
    $c$&  4& $(-2,1)(3,-1)(3,-1)$ & 32 & 112  &  - & - & - & -  \\
\hline \hline
    $D3$& 16 & $(1,0)(1,0)(1,0)$ & \multicolumn{6}{c|} {$\chi_3=2\chi_2=\frac{2}{3}\chi_1=\sqrt{30}$} \\
    $(D7)_3$& 4 & $(0,1)(0,-1)(1,0)$ & \multicolumn{6}{c|} {} \\
\hline
\end{tabular}
\end{center}
\end{table}

\begin{table}
[htb] \footnotesize
\renewcommand{\arraystretch}{1.0}
\caption{D-brane configurations and intersection numbers for $Model-F_3-1$.}
\label{Model-F_3-1}
\begin{center}
\begin{tabular}{|c||c|c||c|c|c|c|c|c|}
\hline
    \multicolumn{2}{|c|}{$Model-F_3-1$} & \multicolumn{7}{|c|}{$[U(4)_C\times U(2)_L\times U(2)_R]_{Observable} \times [Sp(8)]_{Hidden}$}\\
\hline \hline \rm{j} & $N$ & $(n^1,m^1)(n^2,m^2)
(n^3,m^3)$ & $n_{\Ysymm}$& $n_{\Yasymm}$ & $b$  & $b'$ & $c$ & $c'$    \\
\hline \hline
    $a$&  8& $(1,0)(1,1)(1,-1)$ & 0 & 0   & -4 & 2 &  4 & -6 \\
    $b$&  4& $(2,1)(3,-1)(1,0)$ & -2 & 2  &  - & - & 0 & 4  \\
    $c$&  4& $(-2,-1)(2,1)(3,1)$ & -18 & -78  &  - & - & - & -  \\
\hline \hline
    $(D7)_2$& 8 & $(0,1)(1,0)(0,-1)$ & \multicolumn{6}{c|} {$\chi_3=\chi_2={3\over 2}\chi_1=\sqrt{21}$} \\
\hline
\end{tabular}
\end{center}
\end{table}

\begin{table}
[htb] \footnotesize
\renewcommand{\arraystretch}{1.0}
\caption{D-brane configurations and intersection numbers for $Model-F_4-1$.}
\label{Model-F_4-1}
\begin{center}
\begin{tabular}{|c||c|c||c|c|c|c|c|c||c|}
\hline
    \multicolumn{2}{|c|}{$Model-F_4-1$}&\multicolumn{8}{|c|}{$[Sp(16)_C\times U(2)_L\times U(2)_R]_{observable}$}\\
\hline \hline \rm{j} & $N$ & $(n^1,m^1)(n^2,m^2)
(n^3,m^3)$ & $n_{\Ysymm}$& $n_{\Yasymm}$ & $b$  & $b'$ & $c$ &$c'$& K\"ahler moduli    \\
\hline \hline
    $a$&  16& $(1,0)(1,0)(1,0)$ & 0 & 0  &  1 & - & -1 & - & $\chi_1\chi_3=6$ \\
    $b$&  4& $(2,-1)(0,1)(3,-1)$ & -10 & 10  &  - & -& 64 &0&$\frac{\chi_1}{\chi_2}+\frac{12}{\chi_1\chi_2}=1$   \\
    $c$&  4& $(-2,-1)(4,1)(1,1)$ & -6 & -58  &  - & -  & - &-&  \\
\hline
\end{tabular}
\end{center}
\end{table}

\begin{table*}
[htb] \footnotesize
\renewcommand{\arraystretch}{1.0}
\caption{D-brane configurations and intersection numbers for $Model-F_5-1$.}
\label{Model-F_5-1}
\begin{center}
\begin{tabular}{|c||c|c||c|c|c|c|c|}
\hline
    \multicolumn{2}{|c|}{$Model-F_5-1$} & \multicolumn{6}{|c|}{$[U(4)_C\times U(2)_L\times Sp(8)_R]_{Observable} \times [Sp(4)]_{Hidden}$}\\
\hline \hline \rm{j} & $N$ & $(n^1,m^1)(n^2,m^2)
(n^3,m^3)$ & $n_{\Ysymm}$& $n_{\Yasymm}$ & $b$  & $b'$ & $c$    \\
\hline \hline
    $a$&  8& $(-1,-1)(2,1)(2,1)$ & -2 & -30  &  -4 & 0 & 1  \\
    $b$&  4& $(1,0)(1,1)(2,-1)$ & -2 &  2 &  - & - &  0  \\
    $c$&  8& $(1,0)(1,0)(1,0)$ & 0 & 0  &  - & - & -  \\
\hline \hline
    $(D7)_1$& 4& $(1,0)(0,1)(0,-1)$ & \multicolumn{5}{c|} {$\chi_3=2\chi_2$, $\frac{2}{\chi_2^2}+\frac{3}{\chi_1\chi_2}=1$} \\
\hline
\end{tabular}
\end{center}
\end{table*}

\begin{table*}
[htb] \footnotesize
\renewcommand{\arraystretch}{1.0}
\caption{D-brane configurations and intersection numbers for $Model-F_6-1$, here
$\beta^g$ are beta functions for the associated gauge symmetries in the hidden sector.}
\label{Model-F_6-1}
\begin{center}
\begin{tabular}{|c||c|c||c|c|c|c|c|}
\hline
    \multicolumn{2}{|c|}{$Model-F_6-1$}&\multicolumn{6}{|c|}{$[U(4)_C\times Sp(8)_L\times U(2)_R]_{Observable} \times [Sp(4)\times Sp(4)]_{Hidden}$}\\
\hline \hline \rm{j} & $N$ & $(n^1,m^1)(n^2,m^2)
(n^3,m^3)$ & $n_{\Ysymm}$& $n_{\Yasymm}$ & $b$  & $c$ & $c'$   \\
\hline \hline
    $a$&  8& $(1,0)(1,1)(1,-1)$ & 0 & 0   & -1 &  6 & -4  \\
    $b$&  8& $(0,1)(0,-1)(1,0)$ & 0 & 0  &  -  & 3 & -  \\
    $c$&  4& $(-1,-1)(3,1)(2,1)$ & -4 & -44  &  -& - & -  \\
\hline \hline
    $(D7)_1$&  4 & $(1,0)(0,1)(0,-1)$ & \multicolumn{5}{c|} {$\chi_2=\chi_3$, $\frac{6}{\chi_3^2}+\frac{5}{\chi_1\chi_3}=1$} \\
    $(D7)_2$&  4 & $(0,1)(1,0)(0,-1)$ & \multicolumn{5}{c|} {$\beta^g_{(D7)_1}=-3(0)$, $\beta^g_{(D7)_2}=-5(-2)$} \\
\hline
\end{tabular}
\end{center}
\end{table*}

\end{document}